\documentclass[conference]{IEEEtran}
\IEEEoverridecommandlockouts
\usepackage{cite}
\usepackage{amsmath,amssymb,amsfonts}
\usepackage{algorithmic}
\usepackage{graphicx}
\usepackage{textcomp}
\def\BibTeX{{\rm B\kern-.05em{\sc i\kern-.025em b}\kern-.08em
    T\kern-.1667em\lower.7ex\hbox{E}\kern-.125emX}}
\usepackage{amsmath,amsfonts}
\usepackage{algorithmic}
\usepackage{array}
\usepackage[caption=false,font=normalsize,labelfont=sf,textfont=sf]{subfig}
\usepackage{textcomp}
\usepackage{stfloats}
\usepackage{url}
\usepackage{verbatim}
\usepackage{graphicx}
\usepackage{float} 
\usepackage{makecell}
\usepackage{supertabular}
\usepackage[singlelinecheck=false]{caption}
\usepackage{listings}
\usepackage[dvipsnames]{xcolor}
\usepackage{url}
\usepackage{soul}
\usepackage{graphicx}
\usepackage{booktabs}
\usepackage{amssymb}
\usepackage{cite}
\usepackage{hyperref}
\usepackage{academicons}
\colorlet{punct}{red!60!black}
\definecolor{background}{HTML}{EEEEEE}
\definecolor{delim}{RGB}{20,105,176}
\colorlet{numb}{magenta!60!black}
\lstdefinelanguage{json}{
    basicstyle=\scriptsize\ttfamily,
    showstringspaces=false,
    breaklines=true,
    frame=lines,
    backgroundcolor=\color{background},
    literate=
     *{0}{{{\color{numb}0}}}{1}
      {1}{{{\color{numb}1}}}{1}
      {2}{{{\color{numb}2}}}{1}
      {3}{{{\color{numb}3}}}{1}
      {4}{{{\color{numb}4}}}{1}
      {5}{{{\color{numb}5}}}{1}
      {6}{{{\color{numb}6}}}{1}
      {7}{{{\color{numb}7}}}{1}
      {8}{{{\color{numb}8}}}{1}
      {9}{{{\color{numb}9}}}{1}
      {:}{{{\color{punct}{:}}}}{1}
      {,}{{{\color{punct}{,}}}}{1}
      {\{}{{{\color{delim}{\{}}}}{1}
      {\}}{{{\color{delim}{\}}}}}{1}
      {[}{{{\color{delim}{[}}}}{1}
      {]}{{{\color{delim}{]}}}}{1},
}

\begin{document}

\title{Movie Box Office Prediction With Self-Supervised and Visually Grounded Pretraining\\
}

\author{\IEEEauthorblockN{Qin Chao\textsuperscript{1,3}, Eunsoo Kim\textsuperscript{2}, and Boyang Li\textsuperscript{3}}
\IEEEauthorblockA{\textit{\textsuperscript{1}Alibaba Group and
the Alibaba-NTU Joint Research Institute, Singapore} \\
\textit{\textsuperscript{2}Nanyang Business School, Nanyang Technological University (NTU), Singapore} \\
\textit{\textsuperscript{3}School of Computer Science and Engineering, NTU, Singapore}\\
chao0009@e.ntu.edu.sg,  \{eunsoo, boyang.li\}@ntu.edu.sg}
}

\maketitle

\begin{abstract}
Investments in movie production are associated with a high level of risk as movie revenues have long-tailed and bimodal distributions \cite{pan2010}. Accurate prediction of box-office revenue may mitigate the uncertainty and encourage investment. However, learning effective representations for actors, directors, and user-generated content-related keywords remains a challenging open problem. In this work, we investigate the effects of self-supervised pretraining and propose visual grounding of content keywords in objects from movie posters as a pretraining objective. Experiments on a large dataset of 35,794 movies demonstrate significant benefits of self-supervised training and visual grounding. In particular, visual grounding pretraining substantially improves learning on movies with content keywords and achieves 14.5\% relative performance gains compared to a finetuned BERT model with identical architecture. 
\end{abstract}


\section{Introduction}
\label{sec:intro}
Movies are undoubtedly a preeminent form of art in the 21$^\text{st}$-century human civilization. However, the business side of movie production is often less than glamorous. Statistics \cite{pan2010} show that box-office revenues have long-tailed and bimodal distributions, where a small number of movies take most of the profit and the majority barely make even. According to \href{https://www.boxofficemojo.com/year/world/2019/}{boxofficemojo.com}\footnote{\url{https://www.boxofficemojo.com/year/world/2019/}}, in 2019, the top-10 highest-grossing movies collected 13.2 billion US dollars or 37.4\% of the global revenue of the top-200 movies. Three years into the pandemic, as of November 2022, the ratio balloons to 50.1\%. The exorbitant risk of the industry drives producers to focus on superhero movies and sequels, whose outcomes are relatively predictable. Small studios that cannot afford to make high-budget movies share an ever smaller pie. 

Algorithmic box office prediction holds the promise to help producers properly budget expenses, reduce risk, and encourage investment in creative and diverse content. The problem has attracted much research interest \cite{apala2013prediction_lash1, parimi2013prerelease_lash24, simonoff2000predicting_lash26, lash2016earlyprediction_lash, eliashberg2014script1_lash11, rajput2017weibo_sheet11, cizmeci2018predicting_sheet8, shafaei2019starpower1_sheet1, boccardelli2008critical_lash5, kim2023does, quader2017performance_sheet10, antipov2017box_sheet12, zhou2018, kim2019prediction }. In this paper, we investigate the effects of self-supervised pretraining and visual grounding. 

The star power of actors and directors is one of the most important factors determining box office revenue, but the data for each actor and director is limited. Even prolific directors typically make less than 30 movies throughout their careers. Similarly, few modern actors play leading roles in more than 30 movies. By modern machine learning standards, these numbers are considered few-shot settings. To tackle data sparsity, we adopt self-supervised pretraining that encourages the network to learn the data distribution before training on box office data. 

Another important, yet difficult to model, aspect of the box office is the movie content. The movie storyline is a complex artifact with multiple layers of semantics \cite{barot2015tripartite, cutting2016narrative, li2015learning, davis2011distributed}, which are challenging for even state-of-the-art AI to understand. To tackle this issue, we utilize user-generated content keywords from \href{http://www.themoviedb.org}{The Movie Database (TMDB)}\footnote{\url{www.themoviedb.org}} to incorporate the movie content into the box office prediction problem. Table \ref{tab:keywords_result} shows example keywords. Compared to traditional  genre categories, these keywords provide finer-grained categorization of content, including topic, plot, emotion, and even source-related information\footnote{More details are in the keyword contribution guide located at \url{https://www.themoviedb.org/bible/movie}}.

\begin{table}[!t]
    \centering  
    \small
    \caption{Examples of  user-generated keywords from TMDB.}\label{tab:keywords_result}
    \begin{tabular}{@{}c@{}}  
        \toprule
        \makecell{action, criminology, fbi, psycho, aircraft, robot\\ love, hate, high school, father-daughter relationship, \\ paris france,  kingdom, based on novel or book}\\
        \bottomrule
    \end{tabular}
\end{table}

To gain a precise understanding of these keywords, we further propose to ground the keywords in the visual modality --- the movie posters. In the context of movies, the meaning of keywords can differ subtly from their daily usage. For example, the keyword \emph{action} may be associated with explosion, car chasing, or martial art, deviating from its dictionary definition. The keyword \emph{robot} typically refers to robots in science fiction or animation movies, rather than those on assembly lines. Recent research \cite{tan_bansal_2020_vokenization,YujieLu-Imagination-2022} shows that grounding language in visual signals yields improved representation. In this paper, we find that this effect also exists and that the improved representation contributes to a better box office prediction. To our knowledge, this is the first paper that visually grounds textual information for box office prediction.

Overall, our research highlights the effectiveness of self-supervised pretraining and visual grounding in box office prediction. Our models relatively reduce prediction error by 7.8\%$\sim$14.5\% 
compared to the directly finetuned baseline BERT model under the same number of hyperparameters, whereas pretraining with visual grounding leads to up to 2.1\% relative performance improvements.

With this paper, we make the following contributions. First, we propose self-supervised pretraining for movie box office forecasting that can utilize a combination of textual and numerical information. Second, we demonstrate that visual grounding user-generated keywords in movie posters significantly improves pretraining, suggesting a good correlation between movie content and the posters. Finally, we construct a large well-organized \href{https://github.com/jdsannchao/MOVIE-BOX-OFFICE-PREDICTION}{dataset} for movie box office prediction and share it with the research community\footnote{\url{https://github.com/jdsannchao/MOVIE-BOX-OFFICE-PREDICTION}}.

\section{Related Work}
\noindent \textbf{Predicting Movie Success.}    Prior work has attempted to predict a number of indicators of commercial and artistic success, including the box office \cite{ apala2013prediction_lash1, parimi2013prerelease_lash24, simonoff2000predicting_lash26}, return on investment \cite{lash2016earlyprediction_lash, eliashberg2014script1_lash11, rajput2017weibo_sheet11}, IMDb ratings \cite{cizmeci2018predicting_sheet8, shafaei2019starpower1_sheet1}, critic reviews\cite{kim2023does}, and awards or award nominations \cite{boccardelli2008critical_lash5}. Recently, with the advancement of ML, deep networks have begun to gain research attention \cite{quader2017performance_sheet10, antipov2017box_sheet12, zhou2018, kim2019prediction}. 

In terms of features, aside from commonly adopted numeral features, \cite{apala2013prediction_lash1,rajput2017weibo_sheet11,eliashberg2014script1_lash11,lash2016earlyprediction_lash, kim2023does} utilize textual features such as sentiment and topics. In particular, topics from Latent Dirichlet Allocation    \cite{lash2016earlyprediction_lash,kim2023does} may be seen as a type of content feature. \cite{shafaei2019starpower1_sheet1, kim2019prediction} utilize fastText \cite{bojanowski2016enriching} and ELMO \cite{sarzynska2021detecting} word embeddings respectively. To our knowledge, the only prior work using visual features for box office prediction is \cite{zhou2018}, which incorporates movie poster features from a convolutional neural network during training.  In contrast, our work leverages objects inside the poster to visually ground content keywords during pretraining, but do not use the poster during the finetuning stage.

\noindent \textbf{Self-supervised Multimodal Pretraining.} The success of pretrained textual models such as BERT \cite{devlin2018bert} has inspired a series of  pretrained multimodal models \cite{lu2019vilbert, su2019vl, tan2019lxmert, huang2020pixelbert}, often adopting the masked language modeling (MLM) objective. Similar to a denoising autoencoder, the MLM objective trains the model to predict masked portions of the input. This seemingly simple training technique has demonstrated effectiveness across a wide range of downstream applications. 
Another line of work, such as CLIP \cite{radford2021learning_clip} and BLIP \cite{li2022blip}, adopt a pretraining objective that distinguishes between correct image-text pairings and incorrect pairings. 

A classic problem of cognitive science, the symbol grounding problem \cite{Harnad_1990} is concerned with how words can gain their meaning as pointers to other concepts and objects. Computationally grounding textual tokens in visual images has demonstrated success in some applications \cite{tan_bansal_2020_vokenization, YujieLu-Imagination-2022,yang2022z, ZhangLiSai2021, kiros-etal-2018-illustrative,liu-etal-2022-things, long2020generative}. In this work, we use movie posters as a source of visual grounding for the textual tokens --- keywords. A movie poster is a widely used visual medium to advertise a movie long before its release. Thus, we ground the tokens using objects from a single poster, and each token can be related to multiple objects and vice versa. Compared to the aforementioned prior work, which retrieve or generate relevant images for the textual descriptions, in our task correspondences between the keywords and the poster  are not known \emph{a priori} and must be discovered in a multi-instance manner.

\section{Methodology} 
\label{sec:method}
\begin{figure*}[ht]
\centering
\includegraphics[scale = 0.5]{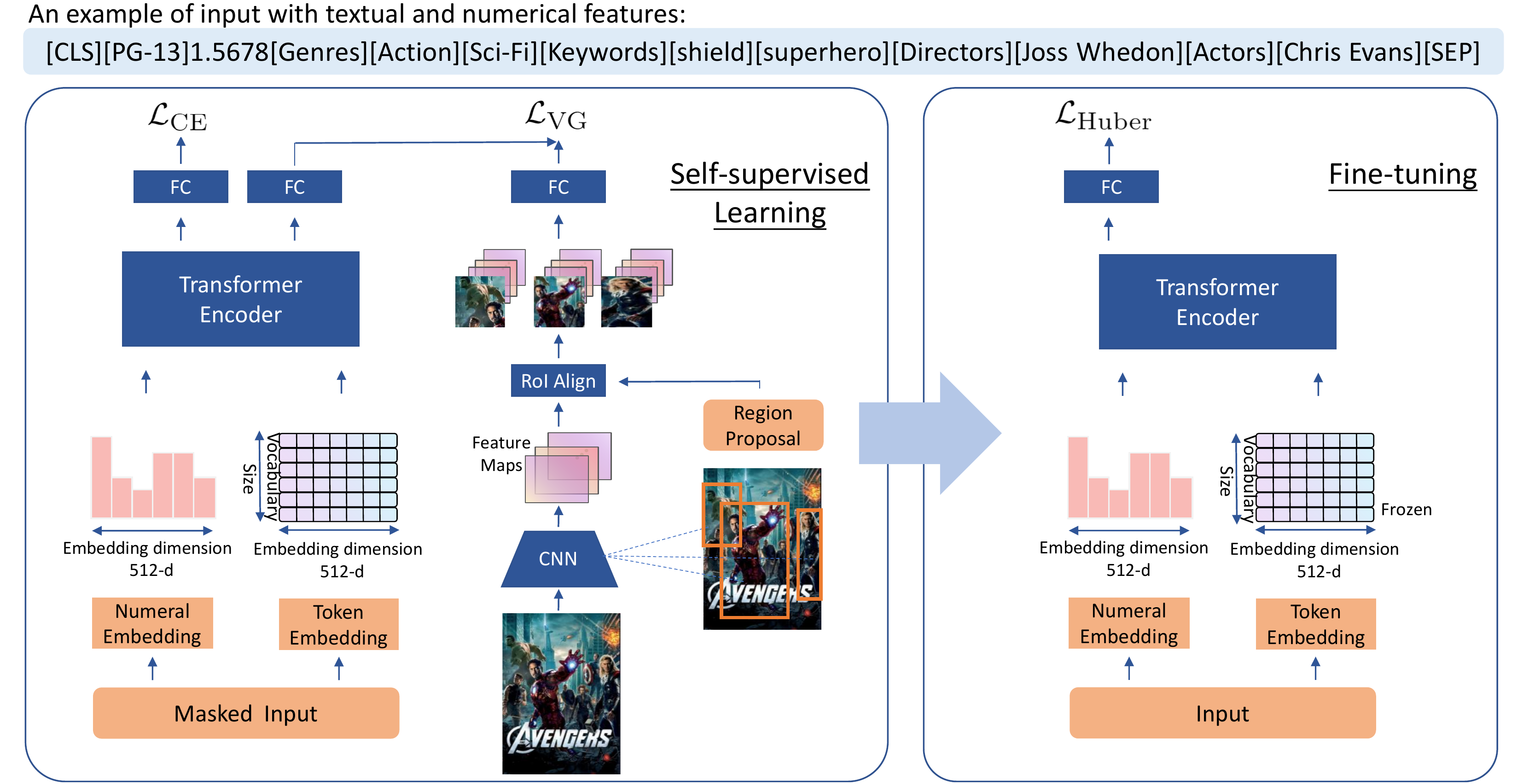}
\caption{The overall pipeline of self-supervised pretraining and finetuning on the box-office prediction task. The token embeddings are frozen during finetuning. }
\label{fig:figures_for_paper_overview}
\end{figure*}

In this section, we first introduce the features used by the proposed network, followed by the pretraining strategies. 

\subsection{Features} \label{section3.1}
We include both discrete features such as actors or directors and real-valued features such as movie budget. 
The embeddings of discrete tokens are learned from data. For real-valued features, we adopt prototype-based numeral embeddings \cite{jin2021numgpt}. Formally, the embedding function is formulated as $\operatorname{NE}(x): \mathbb{R} \rightarrow  \mathbb{R}^{D}$ that maps a real number $x$ to a $D$-dimensional vector with the component 
\begin{equation}
\mathrm{NE}_i(x)=\exp \left(-\frac{\left\|x-q_i\right\|_2}{\sigma^2}\right),  
\end{equation}
where $\{q_i\}^{D-1}_{i=0}$ are $D$ evenly spaced numbers over a specified interval, e.g., $[-10,10]$. Before applying the numeral embedding function, we normalize the values using logarithm or min-max normalization, depending on whether or not the feature has a long-tail distribution. 

We broadly categorize the features used in forecasting box office revenue into four categories: investment \& marketing, star power, content, and competition \& seasonality. 

\noindent \textbf{Investment \& Marketing.} The production budget is often an indicator of the movie's quality. Here we take the logarithm with base 10. Furthermore, we include the distributor company as a token as distributors with greater market power may release movies on more screens, which increases revenue. 

\noindent  \textbf{Star Power.} We include up to two directors, two writers, and three leading actors in our model. Each person is a unique token whose embeddings are trained from scratch. We also calculate the profitability of each person, which is defined as the average of the revenues of all previous movies that this person has participated in as one of the leading roles.
Moreover, we incorporate the gender and age of each actor at the time of the movie release. 

\noindent \textbf{Movie Content.} We first include genres and MPAA ratings. In addition, we also include an indicator for whether a movie is part of a franchise. 

Inspired by the success of user-generated keywords as content descriptors \cite{annalyn2018predicting}, we collect user-generated keywords from TMDB, yielding a total of 7,700 unique keywords for 35,794 movies. Among the keywords, we observe many rare keywords and near-synonyms, which may hinder learning. For rare keywords, the lack of data may prevent accurate embedding estimation. Synonyms and near-synonyms cause problems for constrastive learning, which would force the model to learn dissimilar embeddings for two words with similar meanings. 

To overcome these issues, we cluster the keywords using both lexical similarity and co-occurrence statistics. To capture lexical information, we use 300-dimensional embeddings computed by fastText \cite{bojanowski2016enriching}. Next, we construct a movie-keyword term-frequency inverse-document-frequency (TF-IDF) matrix, which captures the co-occurrence statistics of keywords. From the TF-IDF matrix, we use the technique of \cite{zhang2021initialization} to construct embeddings for keywords. We extract the first 50 dimensions of the singular vectors to represent a keyword. The final representation is the 350-dimensional concatenation of the two vectors. We then perform average-link agglomerative clustering and use the resultant keyword clusters as features of movies. We show detailed cluster results in Appendix \ref{appendix:tables}. 

\noindent  \textbf{Competition \& Seasonality.} To capture the effects of changing consumer tastes and holiday seasons, we include the year and month of the movie release as discrete tokens. Further, we model the competition intensity during the release window. We first identify competitors as those released two weeks before and after the current movie and have the same genre. After that, we sum up the overlap of content keywords, computed using the Jaccard similarity between the current movie and every competitor. 

\subsection{Self-supervised Pretraining }
Figure \ref{fig:figures_for_paper_overview} shows the overall pipeline. In the first stage, we pretrain a Transformer network on the MLM and visual grounding objectives. Next, we freeze the token embeddings and finetune the network on box-office prediction. We now introduce the pretraining tasks. 

\noindent \textbf{Masked Field Prediction.} 
We adopt a pretraining objective similar to the masked language modeling task, which has been shown to be an effective pretraining method for natural language understanding \cite{devlin2018bert} and multimodal understanding \cite{lu2019vilbert}. 
We randomly mask one token from each group of input features: genres, keywords, director/writer names, and actor names. The network is trained to predict the missing token. The prediction is formulated as cross-entropy losses, which we denote as $\mathcal{L}_{CE}$. By training the network to predict missing fields, we encourage the network to learn the correlations between the inputs, which could mitigate data scarcity issues. 

\noindent \textbf{Structured Visual Grounding.} The content of the movie is undoubtedly crucial for box office, but understanding the user-generated content keywords is challenging. In particular, the content keywords may change in the context of motion pictures as the meaning of keywords can differ subtly from their daily usage as mentioned before.

We propose to ground the keywords in the visual modality provided by the movie posters. We conduct contrastive learning that encourages high similarity between a poster and the corresponding content keywords and suppresses the similarity between incorrectly paired posters and keyword sets. We first perform object detection on the poster with an off-the-shelf network, VinVL \cite{zhang2021vinvl}, but our method is not tied to this particular choice. We denote the extracted object features from the $i^{\text{th}}$ movie as $\mathcal{Z}_i = 
\{\boldsymbol{z}_{m}\}^{M}_{m=1}$. Note that we use the subscript $i$ to denote the movie index. 
We also take the contextualized embeddings of the keywords from the output of the Transformer network, denoted as $\mathcal{X}_i ={\{\boldsymbol{x}_{k}\}}^{K}_{k=1}$. 

\begin{figure}[!t]
    \centering
    \includegraphics[scale = 0.46]{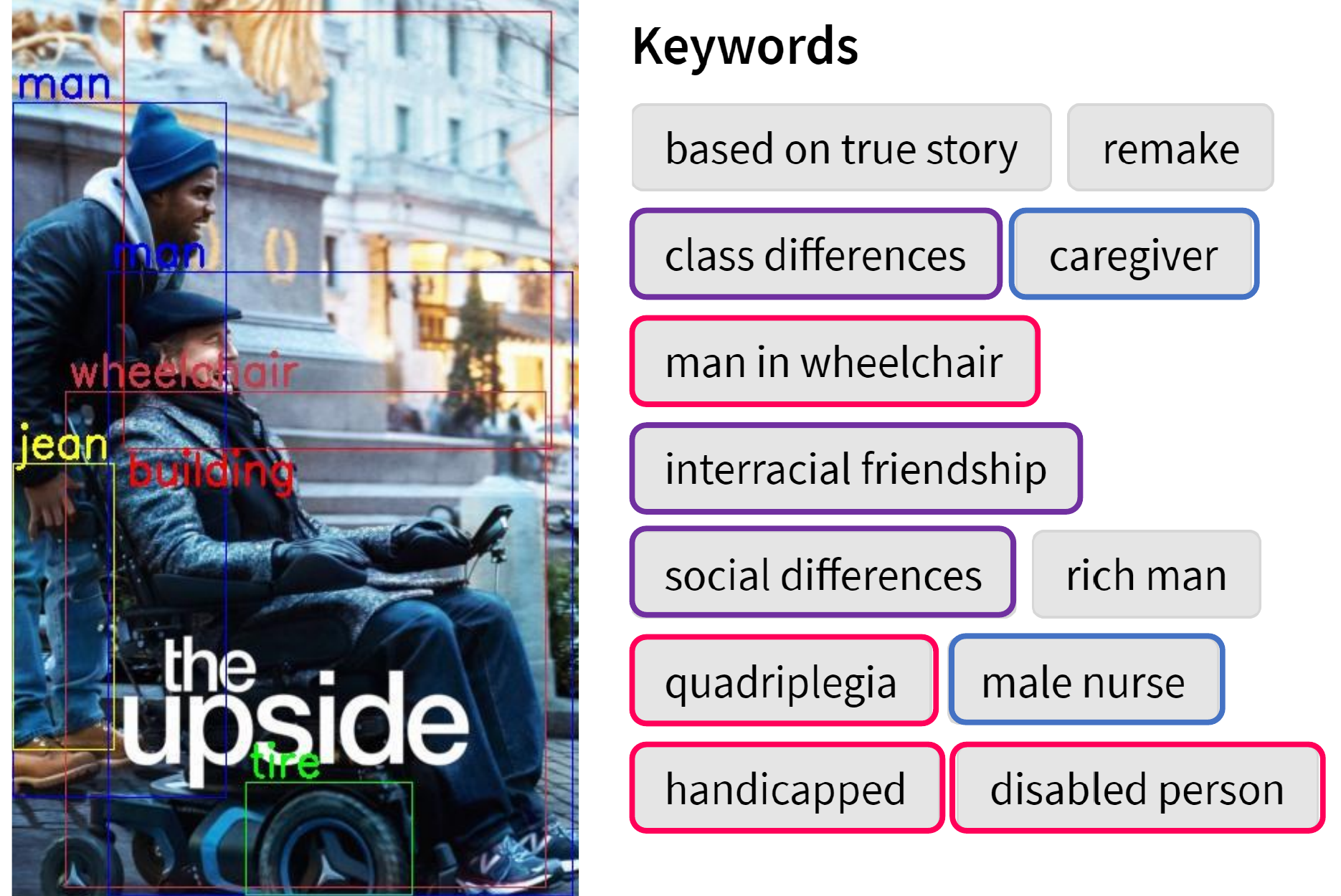}
    \caption{Multiple objects and keywords alignments for the movie \emph{The Upside} (2019)}
    \label{fig:theupside_keywords_links}
\end{figure}

We define the similarity between the poster and the keywords as 
\begin{equation}
\text{sim}(\mathcal{X}_i, \mathcal{Z}_i) = \sum_{(\boldsymbol{x}, \boldsymbol{z})\in \mathcal{X}_i\times \mathcal{Z}_i} \exp(\frac{\boldsymbol{x}^\top \boldsymbol{z}}{\|\boldsymbol{x}\|_2 \|\boldsymbol{z}\|_2}),
\end{equation}
where $\times$ denotes the Cartesian product and $\| \cdot \|_2$ denote the L2 norm. 
To motivate the definition, we show one example poster and the associated keywords in Figure \ref{fig:theupside_keywords_links}. We use colors of the keyword boxes to indicate cluster membership (e.g., ``quadriplegia'' and ``handicapped'' both belong to the red cluster). We observe that a cluster can correspond to multiple objects and one object may ground multiple clusters. For instance, the cluster ``quadriplegia'' is grounded by the wheelchair, the tire and the sitting man; the sitting man relates to the red and the purple clusters. Due to many-to-many relations, we follow \cite{miech2020end} to define the similarity between the two sets as the sum of similarities of all possible pairs. 

With randomly sampled negative pairs $(i^\prime, j^\prime)$, we define the visual grounding loss, $\mathcal{L}_\text{VG}$, as
\begin{equation}
\mathcal{L}_\text{VG} = -\frac{1}{N} \sum_{i=1}^N \log \left(\frac{ \text{sim}(\mathcal{X}_i, \mathcal{Z}_i)}{\text{sim}(\mathcal{X}_i, \mathcal{Z}_i) + \sum_{(i^\prime, j^\prime)} \text{sim}(\mathcal{X}_{i^\prime}, \mathcal{Z}_{j^\prime})}\right)    
\end{equation}
where $N$ is the total number of movies in the training set.

\noindent

\subsection{Finetuning on Box Office Prediction } 
In the finetuning stage, we train the network to predict box office revenues. We generate the prediction by feeding the average output from all input positions to a fully connected layer. Revenues follow a long-tailed distribution, which we approximate using a log-normal distribution. Hence, we take the base-10 logarithm of the revenue as the target value. To further reduce the effects of outliers, we train the network using the smooth L1 loss, also called the Huber loss,
\begin{equation}
\small
\mathcal{L}_\textrm{Huber}= \begin{cases}0.5\left(y-\hat{y}\right)^2,     \text{ if}\left|y-\hat{y}\right|< 1  \\ 
\left|y-\hat{y}\right|-0.5  , \text{ otherwise}\end{cases},
\end{equation}
where $y$ is the ground truth and $\hat{y}$ is the prediction.

\section{Experimental Results}
\subsection{Data and Experimental Setup }
We collect metadata of 35,794 movies from TMDB, including the period from 1920 to 2020. Total box office data for each movie during its original release period is crawled from IMDbPro. We use stratified sampling to divide the data into train, validation, and test sets in the ratios of 70/10/20, using ``franchise movie'' as the label for stratification. Using the method in \S \ref{section3.1}, we cluster 7,700 keywords into 1,414 clusters. The number of clusters is tuned on the validation set.
We use a 4-layer Transformer, with model dimension $d_{\text{model}}=512$, fully connected layer dimension $d_{ff}=512$, and 4 attention heads. The architecture is the same as $\operatorname{BERT}_\text{small}$. More hyperparameters are reported in the Appendix  \ref{appendix:parameters}.

\subsection{Baselines } 
We introduce three types of baseline models. The first is a Random Forest (RF). We feed only numerical features to the RF as one-hot encodings of the discrete features would have too many dimensions. Next, we introduce pretrained BERT models of small and medium sizes and finetune them on box office prediction. To mimic the classic BERT input, we concatenate all the input tokens into one sentence, while rounding numeral features to one decimal point, and then apply the BERT tokenizer. Lastly, we compare against a random initialized $\operatorname{BERT}_\text{small}$ directly trained on box office prediction and a $\operatorname{BERT}_\text{small}$ with pretrained $\operatorname{BERT}$ embeddings for actors, crew members, genres and keywords. When a name contains multiple words, we use the average of the pretrained $\operatorname{BERT}$ embeddings. For a keyword cluster, we use the embedding of the keyword in the cluster appearing the most frequently. 


\begin{table}[!t]
    \centering  
    \small
    \caption{Performance comparisons on the held-out box office test dataset. Our best model shows a 14.5\% of accuracy improvement compared to $\operatorname{BERT}_\text{small}$.}\label{tab:main_result}
\begin{tabular}{@{}lll@{}}
\toprule        
Model & \multicolumn{2}{c}{Test Huber Loss$_\text{(\% improvement)}$}  \\ 
\midrule
\textbf{Numerical features only} & & \\
Random Forest & \multicolumn{2}{l}{$0.3677_{\textcolor{red}{(-3.5\%)}}$}\\
\midrule
\textbf{Textual and numerical features}\\
$\operatorname{BERT}_\text{small}$ finetuned &$0.3553_\text{(baseline)}$ & \\ 
$\operatorname{BERT}_\text{medium}$ finetuned & $0.3446_{(2.5\%)}$ &\\ 
\midrule
\textbf{Our models} &\makecell[l] {Clustering} &\makecell[l]{Keywords} \\
\midrule
Random init.    & $0.3290_{(7.4\%)}$ & $0.3265_{(8.1\%)}$ \\
\text{\, +} MLM pretraining & $0.3109_{(12.5\%)}$ & $0.3133_{(11.8\%)}$ \\
\text{\, +} VG pretraining & $0.3070_{(13.6\%)}$ & $0.3109_{(12.5\%)}$ \\ 
\midrule        
$\operatorname{BERT}$ embeddings init.   & $0.3137_{(11.7\%)}$ & $0.3249_{(8.6\%)}$ \\        
\text{\, +} MLM pretraining & $0.3102_{(12.7\%)}$ & $0.3226_{(9.2\%)}$ \\
\text{\, +} VG pretraining & $0.3037_{(14.5\%)}$ & $0.3182_{(10.4\%)}$ \\
\bottomrule
\end{tabular}
\end{table}

\subsection{Results and Discussion }
In Table \ref{tab:main_result}, we report the test-set Huber loss for all models, as well as their performance relative to $\operatorname{BERT}_\text{small}$. Pretrained $\operatorname{BERT}$ models easily outperform the RF baseline, but are inferior to the MLM and VG pretraining. Although the larger $\operatorname{BERT}_\text{medium}$ outperforms $\operatorname{BERT}_\text{small}$, it underperforms our MLM-pretrained networks by more than 10\% relatively. The domain gap between movie and textual data used in pretraining and our feature engineering likely contribute to the performance gaps.


Notably, VG pretraining obtains sizeable improvements on top of MLM for both types of embedding initialization. The fact that VG pretraining leads to improvement even with BERT-pretrained token embeddings corroborates our hypothesis that keywords may have specialized meanings in the movie context and visual grounding may help capture the specialized semantics. Finally, the best test loss of 0.3037, or 14.5\% improvement relative to $\operatorname{BERT}_\text{small}$, is achieved by MLM+VG pretraining. 

\begin{figure}[!t]
    \centering
    \includegraphics[scale = 0.53]{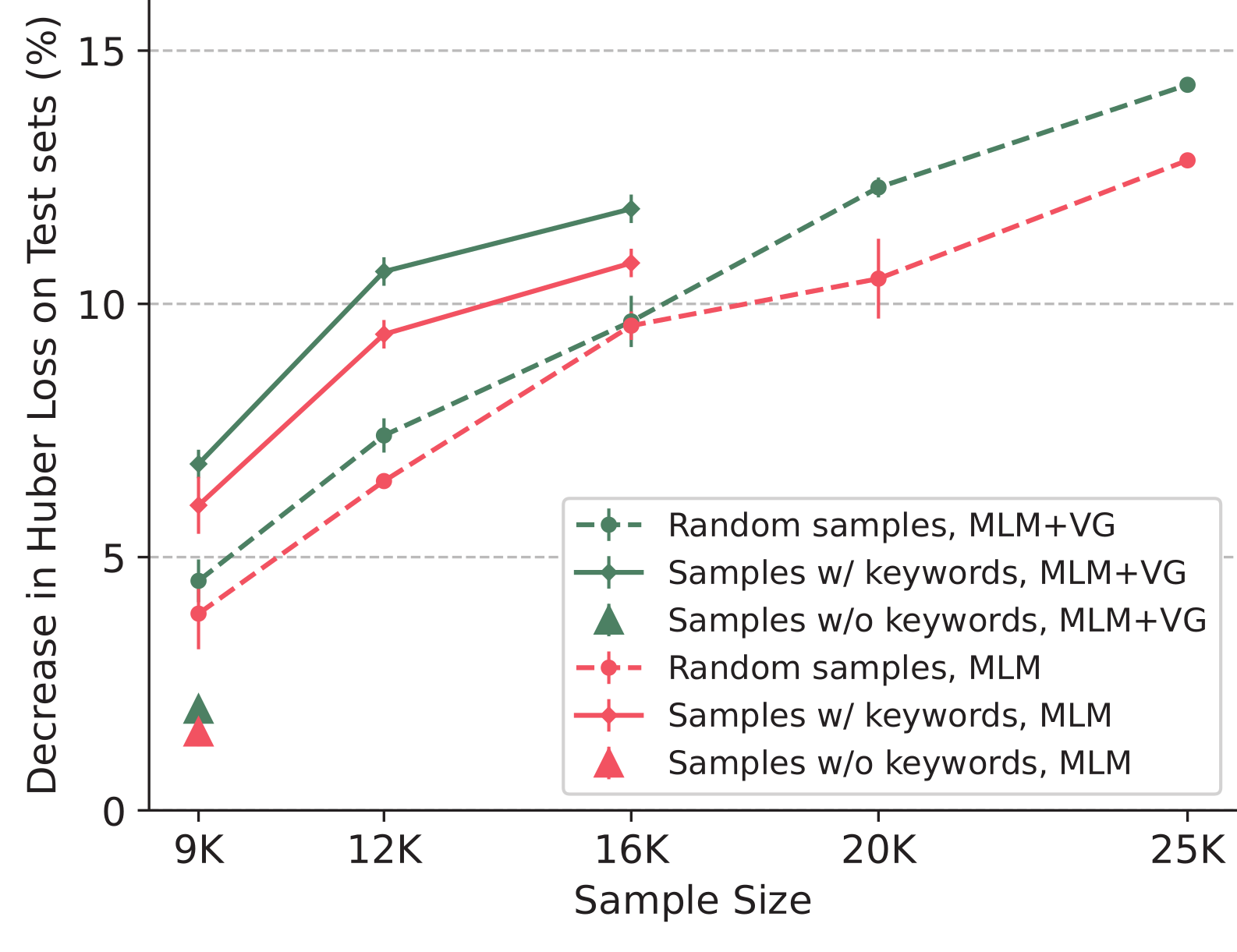}
    \caption{Test losses on box office prediction with different training set sizes. Vertical bars indicate standard deviations. Exact numbers are reported in Appendix \ref{appendix:adjustlossweights}.}
    \label{fig:figure2}
\end{figure}

\noindent \textbf{Content Keywords and Scaling.}
As not all movies come with user-supplied keywords, we further investigate the effects of pretraining on movies with and without content keywords. We split the training set into movies with keywords (16K out of 25K) and movies without (9K out of 25K). As comparison baselines, we also create random subsets of the entire training set of sizes 9K, 12K, 16K, 20K, and 25K. We report losses on the same test set when the MLM+VG network is training on different training sets in Fig. \ref{fig:figure2}. We note that with equal amount of training data, MLM and VG both exhibit stronger generalization when training on data with keywords than randomly mixed data. This agrees with our intuition as MLM exploits correlation between keywords and VG further reinforces keywords with visual information. In Fig. \ref{fig:figure_for_report_appendix_tb5} in the Appendix, we examine if VG improves upon MLM for movies with keywords. We observe that the improvement of MLM+VG over MLM widens as training data increase, suggesting VG scales well and its effectiveness grows with data. 

\begin{figure}[t]
    \centering
    \includegraphics[scale = 0.39]{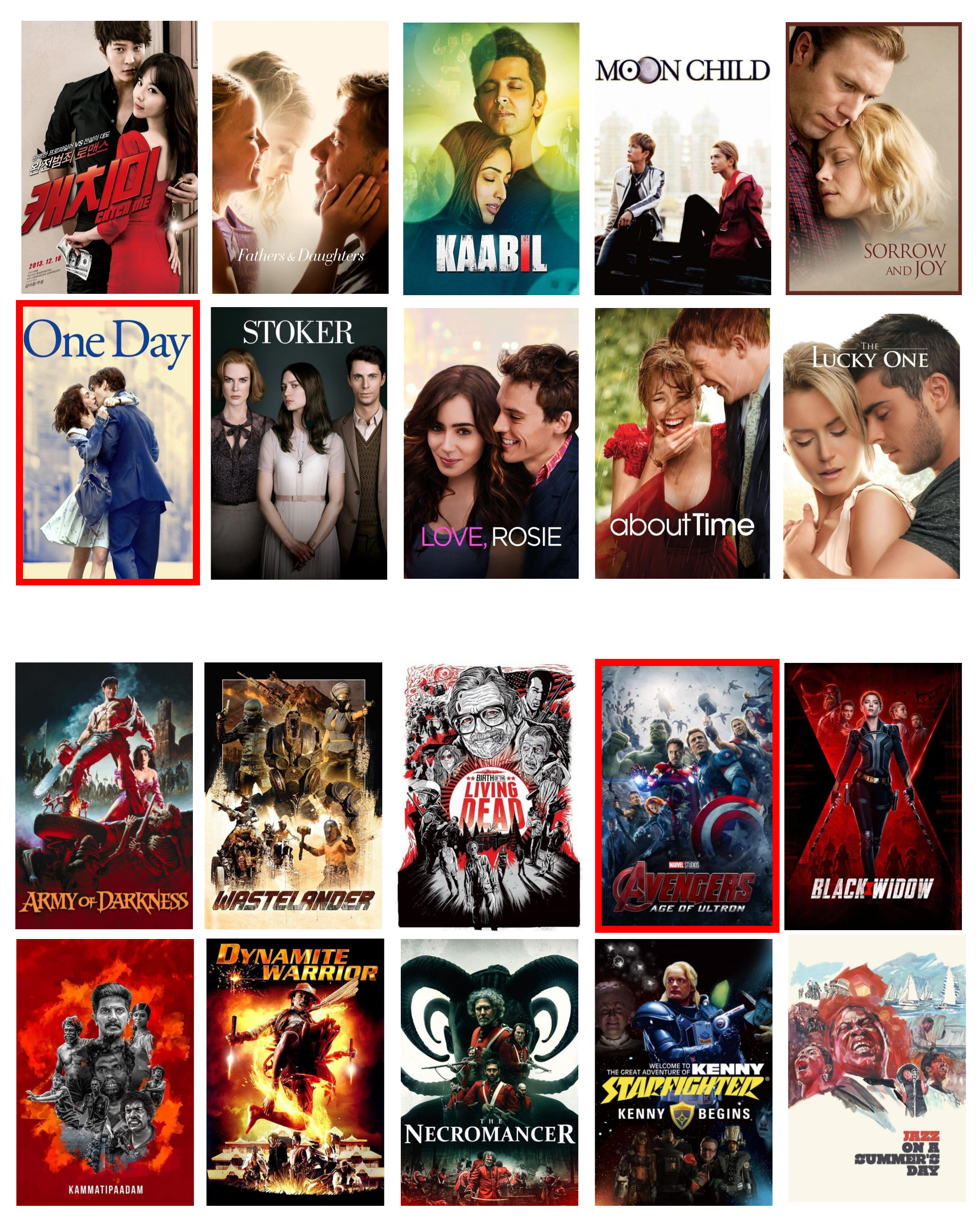}
    \caption{\textbf{Top}: Retrieved posters from the keyword ``love'' in the context of a romantic movie, \emph{One Day (2009)}; \textbf{Bottom}: Retrieved posters using the keyword ``superhero'' in the context of \emph{The Avengers (2012)}.}  \label{fig:posterretrieval_romance_superhero}
\end{figure}


\noindent \textbf{Effects of Keywords Clustering.}
We examine the effects of keyword clusters. Table \ref{tab:main_result} compares results with keyword clustering ("Clustering") with those on raw keywords ("Keywords"). In most cases, keyword clusters provide performance gains, especially when pretrained $\operatorname{BERT}$ embeddings are used. A possible reason is that near-synonyms have similar $\operatorname{BERT}$ embeddings that are difficult for the model to differentiate, and clusters alleviate this problem.

\subsection{Poster Retrieval Examples}
We qualitatively examine the effects of visual grounding. Figure \ref{fig:posterretrieval_romance_superhero} shows posters that are most similar to keywords within the contexts of movies. The top two rows are retrieved for the keyword ``love'' in the context of a romantic movie \emph{One Day (2009)}. The majority of posters fall under the romance genre and visualize a couple embracing one another.
The bottom ten posters are retrieved for the keyword ``superhero'' in the movie \emph{The Avengers (2012)}. The results are mostly action movies with a hero at the center of the poster surrounded by others. Appendix \ref{appendix:otherposterretrievalresults} contains more examples.

\section{Conclusion}
Box office revenue is influenced by a plethora of entangled factors that are often hard to observe, let alone computationally model.  An important challenge in box office prediction is hence to learn representations that capture movie semantics and correlate well with the target variable. 
For this purpose, we propose to pretrain a transformer network with masked language modeling and visual grounding objectives, which demonstrate substantial performance boost. We hope these results could inspire subsequent research on multimodal box-office prediction.

\section{Acknowledgments}
This research is supported, in part, by Alibaba Group through the Alibaba Innovative Research (AIR) Program and Alibaba-NTU Singapore Joint Research Institute (No. AN-GC-2020-011), the National Research Foundation Fellowship (No. NRF-NRFF13-2021-0006), Singapore,  and NTU Start-Up Grant.

\vspace{12pt}

\newpage

\appendix
\subsection{Experimental Setup and Hyperparameters}\label{appendix:parameters}

During the visual grounding pretraining, we randomly select up to 6 keywords per movie and up to 20 objects per poster to compute the similarity. The feature maps of each object have a dimension of $2048\times4\times4$, it is the output from VinVL after the ROI Align \cite{he2017maskrcnn} and an adaptive average pooling layer. After that, the feature map is then flattened spatially and linearly projected to $\mathrm{R}^ {d_{model}}$, where $d_{model} = 512$. 

We use a batch size of 2048 when pretraining the model under MLM objective and reduce the size to 326 when adding the VG objective. The learning rate is 3e-4. The optimizer we used is Adam with weight decay equals to 1e-4. During the fine tune stage, we search for the best performance on the validation dataset in the combinations of learning rate in [1e-3, 3e-4, 1e-4] and batch size in [328, 512, 1024].
\vspace{2mm}
\subsection{Clustering Samples and the Number of Unique Tokens}\label{appendix:tables}
\begin{table}[htbp]
    \centering  
    \small
    \caption{Some examples of the clustering results. The representative word of a cluster is the most frequent keyword in this set.}\label{tab:clusters_result}
    \begin{tabular}{|c|c|}  
        \hline
        Cluster & Elements \\
        \hline
        love & \makecell[l]{'love', 'loved', 'hate', 'unhappy',\\'waiting', 'happy', 'grateful', 'lucky',\\'expecting', 'loving'}\\
        \hline
        superhero & \makecell[l]{'superhero', 'villainess', 'villain', 'symbiote',\\ 'sidekick', 'superhuman', 'teamup', 'nemesis',\\'superheroes', 'supervillain'}\\
        \hline
        psycho & \makecell[l]{'psycho', 'psychotic', 'pyromaniac',\\ 'psychopathic','homicidal', 'deranged'}\\
        \hline        
    \end{tabular}

\end{table}
\begin{table}[htbp]
    \centering  
    \small
    \caption{\# unique tokens for each textual features}\label{tab:unique_tokens}
    \begin{tabular}{|l|l|l|}  
        \hline
        Feature Name & Example & \makecell[l]{No. Unique\\Tokens}\\
        \hline
        Release Year & \makecell[l]{release year range \\from 1920-2020} & 100\\
        \hline
        Release Month & 12 tokens for 12 months & 12\\
        \hline
        MPAA & \makecell[l]{G, PG, PG-13, R, NC17,\\ NotRated, N.A.} & 7 \\
        \hline
        \makecell[l]{Production\\Company} & \makecell[l]{group small studios \\(produced less than 10 movie) \\into `Others'}  & 594  \\
        \hline        
        \makecell[l]{Distributor\\Company} & \makecell[l]{group small studios \\(produced less than 10 movie) \\into `Others'}  & 407  \\
        \hline        
        Franchise & yes or no & 2 \\
        \hline        
        Copycat & yes or no  & 2 \\
        \hline        
        Genres & e.g. Drama, Romance & 18 \\
        \hline        
        Keywords & e.g. `friendship' & 1414 \\
        \hline        
        Crew Names &  \makecell[l]{e.g. `Steven Spielberg'\\is a token} & 15333 \\
        \hline        
        Cast Names &  \makecell[l]{e.g. `Leonardo DiCaprio'\\is a token} & 20366 \\
        \hline
    \end{tabular}
\end{table}
\noindent

\subsection{Meta Data From TMDB}\label{appendix:rawdata}
\begin{lstlisting}[language=json,firstnumber=1]
{'adult': False,
 'backdrop_path': 'c1BaOxC8bo5ACFYkYYxL0bBWRaq.jpg',
 'belongs_to_collection': None,
 'budget': 4000000,
 'genres': [{'id': 80, 'name': 'Crime'}, {'id': 35, 'name': 'Comedy'}],
 'homepage': 'https://www.miramax.com/movie/four-rooms/',
 'id': 5,
 'imdb_id': 'tt0113101',
 'original_language': 'en',
 'original_title': 'Four Rooms',
 'overview': "It's Ted the Bellhop's first night on the job...and the hotel's very unusual guests are about to place him in some outrageous predicaments. It seems that this evening's room service is serving up one unbelievable happening after another.",
 'popularity': 15.811,
 'poster_path': '75aHn1NOYXh4M7L5shoeQ6NGykP.jpg',
 'production_companies': [{'id': 14,
   'logo_path': 'm6AHu84oZQxvq7n1rsvMNJIAsMu.png',
   'name': 'Miramax',
   'origin_country': 'US'},
  {'id': 59,
   'logo_path': 'yH7OMeSxhfP0AVM6iT0rsF3F4ZC.png',
   'name': 'A Band Apart',
   'origin_country': 'US'}],
 'production_countries': [{'iso_3166_1': 'US',
   'name': 'United States of America'}],
 'release_date': '1995-12-09',
 'revenue': 4257354,
 'runtime': 98,
 'spoken_languages': [{'english_name': 'English',
   'iso_639_1': 'en',
   'name': 'English'}],
 'status': 'Released',
 'tagline': "Twelve outrageous guests. Four scandalous requests. And one lone bellhop, in his first day on the job, who's in for the wildest New year's Eve of his life.",
 'title': 'Four Rooms',
 'video': False,
 'vote_average': 5.7,
 'vote_count': 2146}

{'id': 5,
 'keywords': [{'id': 612, 'name': 'hotel'},
  {'id': 613, 'name': "new year's eve"},
  {'id': 616, 'name': 'witch'},
  {'id': 622, 'name': 'bet'},
  {'id': 922, 'name': 'hotel room'},
  {'id': 2700, 'name': 'sperm'},
  {'id': 9706, 'name': 'anthology'},
  {'id': 12670, 'name': 'los angeles, california'},
  {'id': 160488, 'name': 'hoodlum'},
  {'id': 187056, 'name': 'woman director'}]}

{'id': 5,
 'cast': [{'adult': False,
   'gender': 2,
   'id': 3129,
   'known_for_department': 'Acting',
   'name': 'Tim Roth',
   'original_name': 'Tim Roth',
   'popularity': 15.779,
   'profile_path': '/qSizF2i9gz6c6DbAC5RoIq8sVqX.jpg',
   'cast_id': 42,
   'character': 'Ted the Bellhop',
   'credit_id': '52fe420dc3a36847f80001b7',
   'order': 0},
   ...
  'crew': [{'adult': False,
   'gender': 1,
   'id': 3110,
   'known_for_department': 'Directing',
   'name': 'Allison Anders',
   'original_name': 'Allison Anders',
   'popularity': 0.6,
   'profile_path': '/ln8nIx6UjxpMLVQlStCJpx6fyL7.jpg',
   'credit_id': '52fe420dc3a36847f800012d',
   'department': 'Directing',
   'job': 'Director'},
    {'adult': False,
   'gender': 1,
   'id': 3110,
   'known_for_department': 'Directing',
   'name': 'Allison Anders',
   'original_name': 'Allison Anders',
   'popularity': 0.6,
   'profile_path': '/ln8nIx6UjxpMLVQlStCJpx6fyL7.jpg',
   'credit_id': '52fe420dc3a36847f80001c9',
   'department': 'Writing',
   'job': 'Writer'}
\end{lstlisting}

\subsection{Model Input}\label{appendix:inputdata}

\begin{table}[H]
\begin{tabular}{ll}
\textbf{Feature   Name} & \textbf{Tokens / Values}        \\
\makecell[l]{release\_year \\((is the input of both RF model and\\ Transformer model))}           & 2012                            \\
release\_month (both)          & March                           \\
MPAA (both)                     & PG-13                           \\
Budgets (both)                & 7.892094608                     \\
producer (both)                 & Lionsgate                       \\
distributor (both)              & Lionsgate                       \\
N\_competiters (both)           & 0.693147181                     \\
competiter\_similarity (both)   & 0                               \\
franchise (both)                & Yes                             \\
collection name       & The Hunger Games\_0             \\
N\_person (both)                & 0.693147181                     \\
N\_man (both)                   & 0                               \\
N\_woman (both)                & 0                               \\
                        & {[}genres{]}                    \\
genres (both)                   & Adventure                       \\
                        & Fantasy                         \\
                        & Science Fiction                 \\
clusters                & {[}clusters{]}                  \\
                        & retelling                       \\
                        & socialism                       \\
                        & backgammon                      \\
                        & interpretation                  \\
                        & {[}Directors{]}                 \\
Directors               & Gary Ross                       \\
Director1 experience (both)    & 1.098612289                     \\
Director1 profitability (both)  & 0.875335786                     \\
same for Director2      &                                 \\
Writers                 & {[}Writers{]}                   \\
                        & Billy Ray                       \\
Writer1 experience (both)       & 1.386294361                     \\
Writer1 profitability (both)    & 0.86888262                      \\
similar for Writer2     & Gary Ross                       \\
                        & 0                               \\
                        & 0                               \\
Actors                  & {[}Actors{]}                    \\
Actor1                  & Jennifer Lawrence               \\
Actor1 Gender           & Female                          \\
Actor1 Age              & 22                              \\
Actor1 experience (both)      & 0.693147181                     \\
Actor1 profitability (both)     & 0.792347251                     \\
same for Actor2, Actor3 & Josh Hutcherson                 \\
                        & Male                            \\
                        & 20                              \\
                        & 1.098612289                     \\
                        & 0.928510042                     \\
                        & Liam Hemsworth                  \\
                        & Male                            \\
                        & 22                              \\
                        & 0                               \\
                        & 0                              
\end{tabular}
\end{table}

\subsection{More Experimental results}\label{appendix:adjustlossweights}

\begin{figure}[t]
    \centering
    \includegraphics[scale = 0.1]{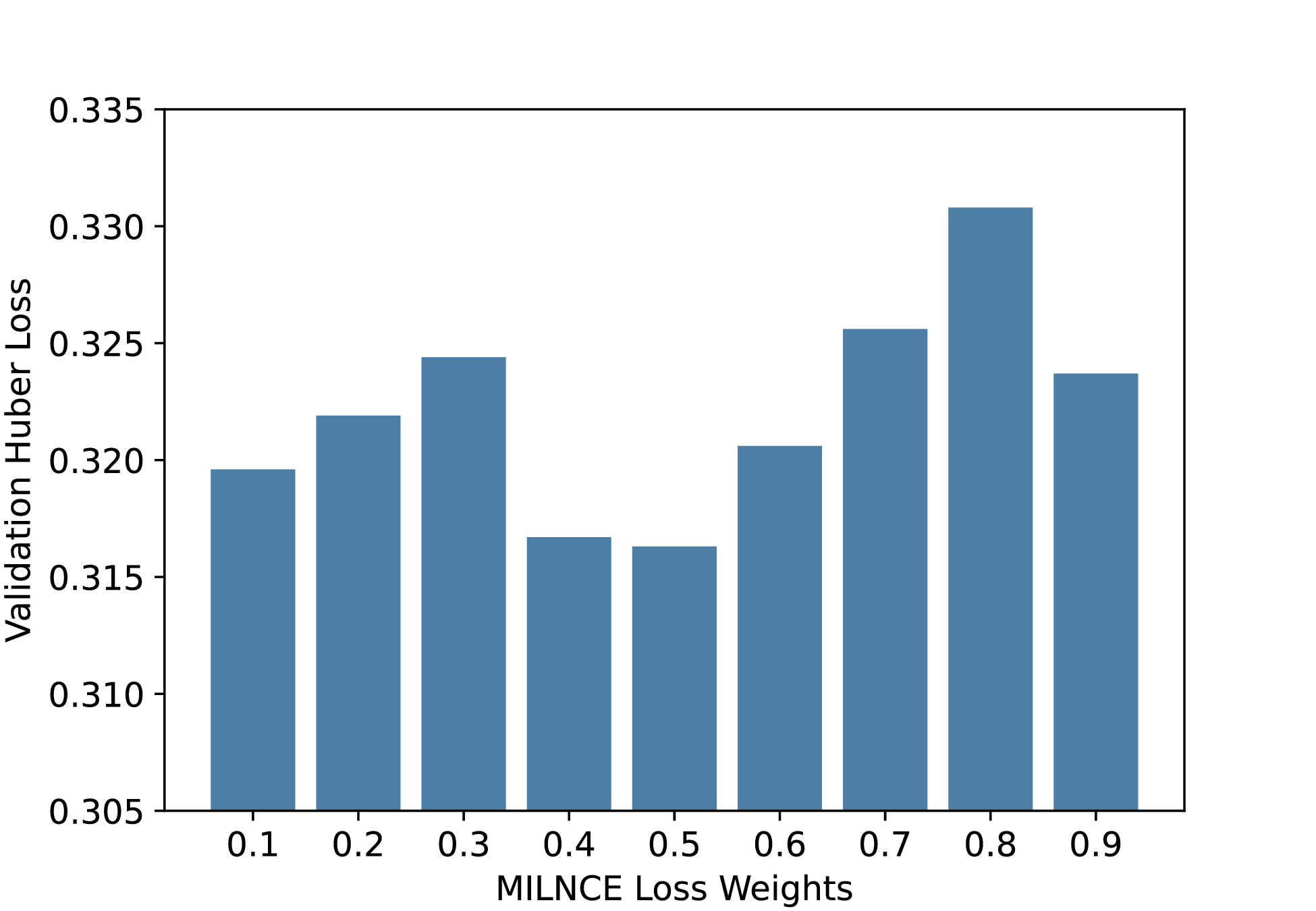}
    \caption{Analysis on VG loss weights when pretrain the model on MLM and VG objective jointly.}
    \label{fig:figure_for_report_fig_2_v3}
\end{figure}

\noindent \textbf{Adjust Loss Weights.} Figure \ref{fig:figure_for_report_fig_2_v3}  shows the impact of the VG loss weight during pretraining on test Huber loss in the fine-tuning stage. That is, we vary the weight of VG loss and MLM loss to examine its impact on the test Huber loss. Based on our experiment, setting the equal weight to the MLM and VG loss attains the lowest test Huber loss.
The results suggest that the movie context information and visual grounding equally contribute to the keyword representations, showcasing the non-negligible impact of visual grounding. 

\noindent \textbf{Detailed Comparison between MLM and MLM+VG.} In Table \ref{tab:appendix_tb5}, we show detailed numbers comparing the two pretraining objectives, MLM and MLM+VG. For each model, we run three random trials and then compute the average and standard deviation. These numbers are reflected in Fig. \ref{fig:figure2} in the main text. 

\begin{figure}[t]
    \centering
    \includegraphics[scale = 0.1]{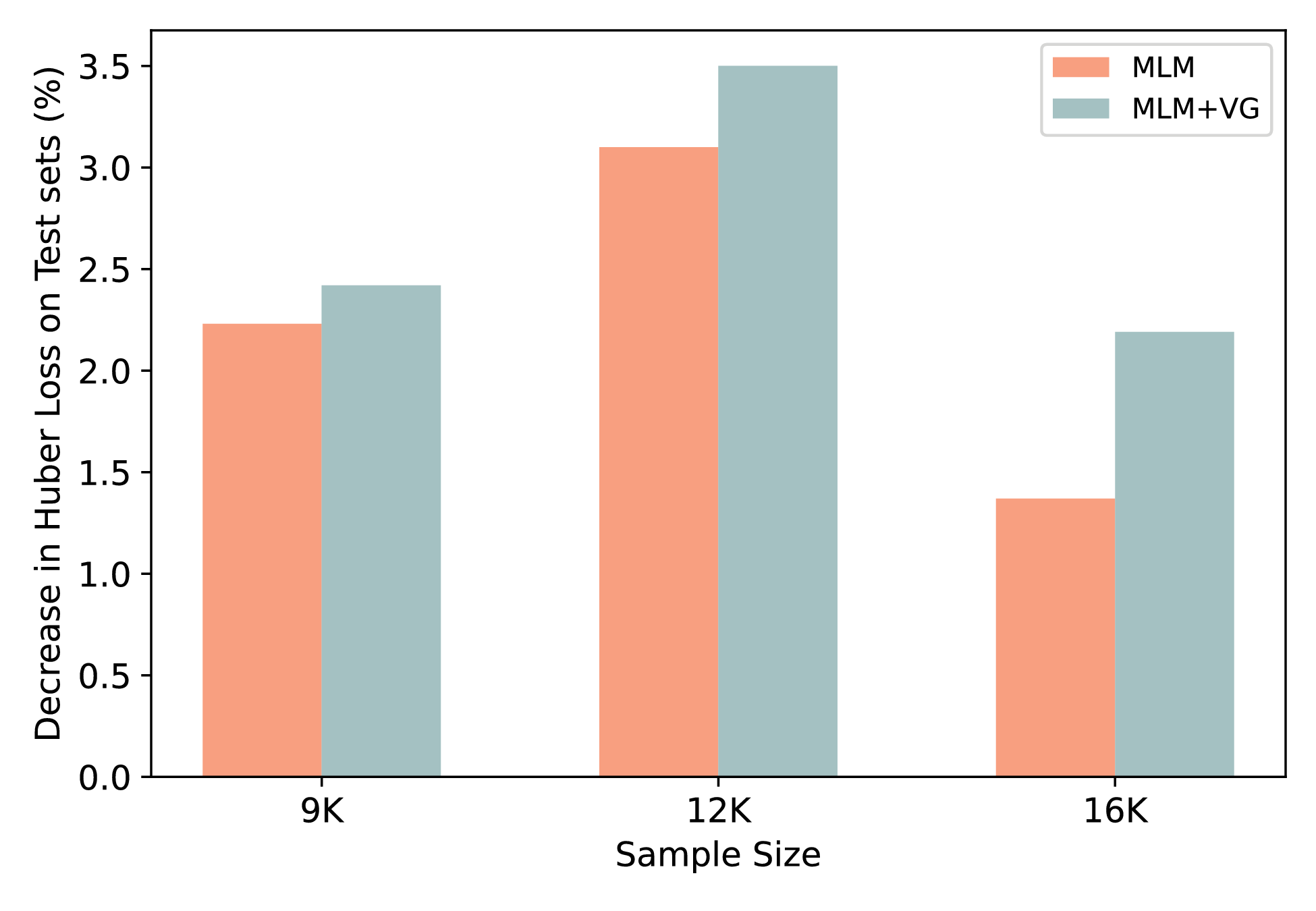}
    \caption{Huber loss decrease between training on data with keywords and random data, for both MLM pretraining model and MLM+VG pertaining model with sample size vary. The gap between MLM and MLM+VG grows as the sample size increases.}
    \label{fig:figure_for_report_appendix_tb5}
\end{figure}

In Fig. \ref{fig:figure_for_report_appendix_tb5}, we show relative test loss improvement of training on data with keywords over training on randomly sampled data, which contain movies with keywords and without keywords. The green bars indicate the average improvements of the MLM+VG model and the red bars indicate the average improvements of the MLM model. We note that as the sample size increases, the gap between the MLM+VG model and the MLM-only models increases, suggesting that the effectiveness of visual grounding increases with data.

\begin{table}[h]
\centering
\small
\caption{ For each model, we run three random trials and then compute the average and standard deviation}\label{tab:appendix_tb5}
\begin{tabular}{cccc} 
\toprule
& & \multicolumn{2}{c}{Average Test Huber Loss(std.)}  \\
\cline{3-4}
& sample size & \makecell[c]{MLM \\ pretraining}  & \makecell[c]{MLM+VG \\ pretraining} \\
\midrule
random samples & 25k         & $0.3097_{(0.0004)}$       & $0.3044_{(0.0005)}$       \\ 
               & 20k         & $0.3180_{(0.0028)}$       & $0.3116_{(0.0007)}$        \\
               & 16k         & $0.3213_{(0.0010)}$       & $0.3201_{(0.0018)}$       \\
               & 12k         & $0.3322_{(0.0002)}$       & $0.3290_{(0.0012)}$       \\
               & 9k          & $0.3415_{(0.0025)}$       & $0.3392_{(0.0015)}$       \\ \midrule
with keywords  & 16k         & $0.3169_{(0.0012)}$       & $0.3131_{(0.0009)}$        \\
               & 12k         & $0.3219_{(0.0007)}$       & $0.3175_{(0.0011)}$         \\
               & 9k          & $0.3339_{(0.0020)}$       & $0.3310_{(0.0012)}$         \\ \midrule
w/o keywords   & 9k          & $0.3498_{(0.0014)}$       & $0.3482_{(0.0002)}$       \\ \bottomrule
\end{tabular}
\end{table}
\pagebreak
\section{More Poster Retrieval Results}\label{appendix:otherposterretrievalresults}

\begin{figure}[h]
    \centering
    \includegraphics[scale = 0.35]{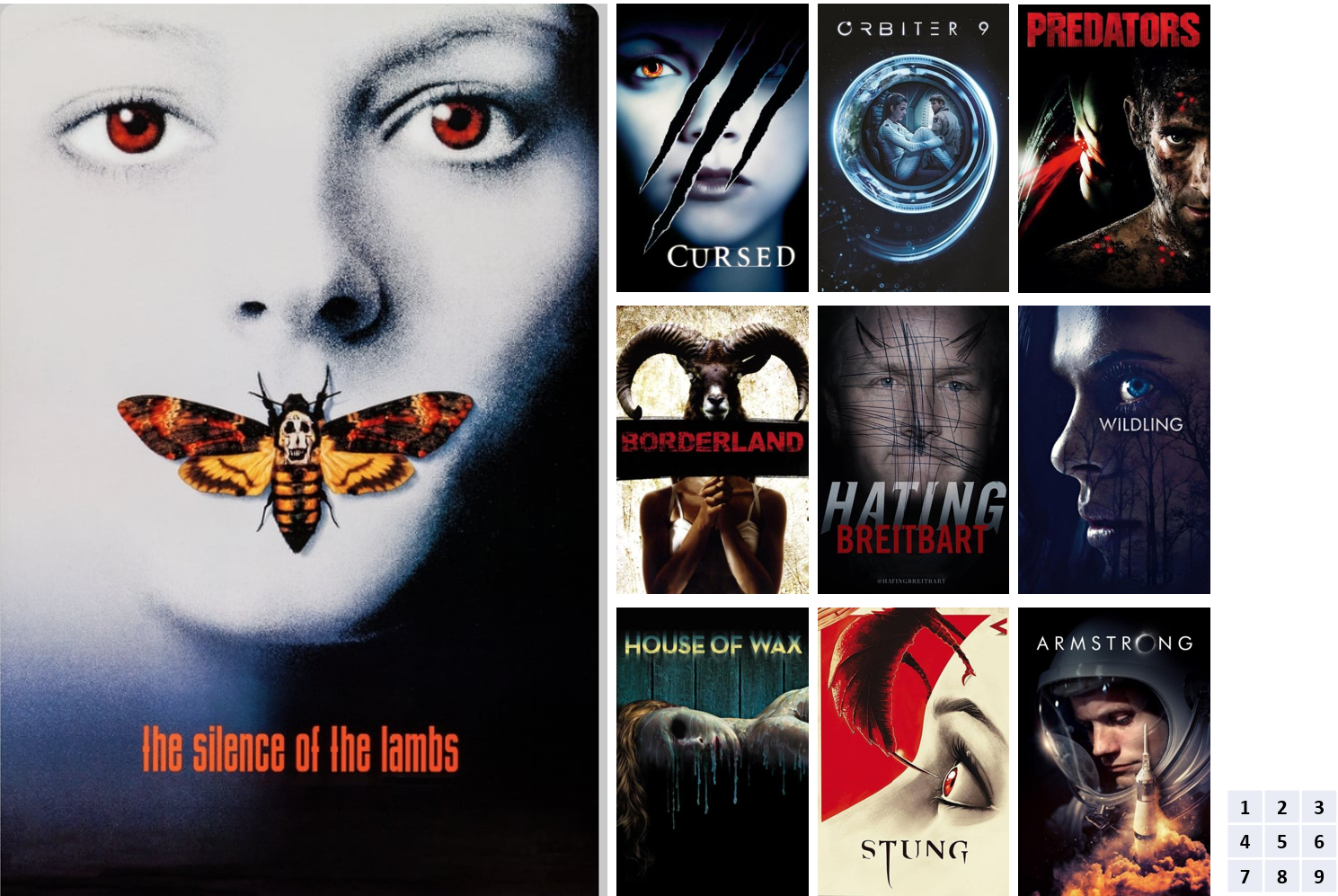}
    \caption{Poster retrieval using the keyword `psycho' in the context of a typical Thriller movie \emph{The Silence of the Limb (1991)}}
\end{figure}

\begin{figure}[H]
    \centering
    \includegraphics[scale = 0.35]{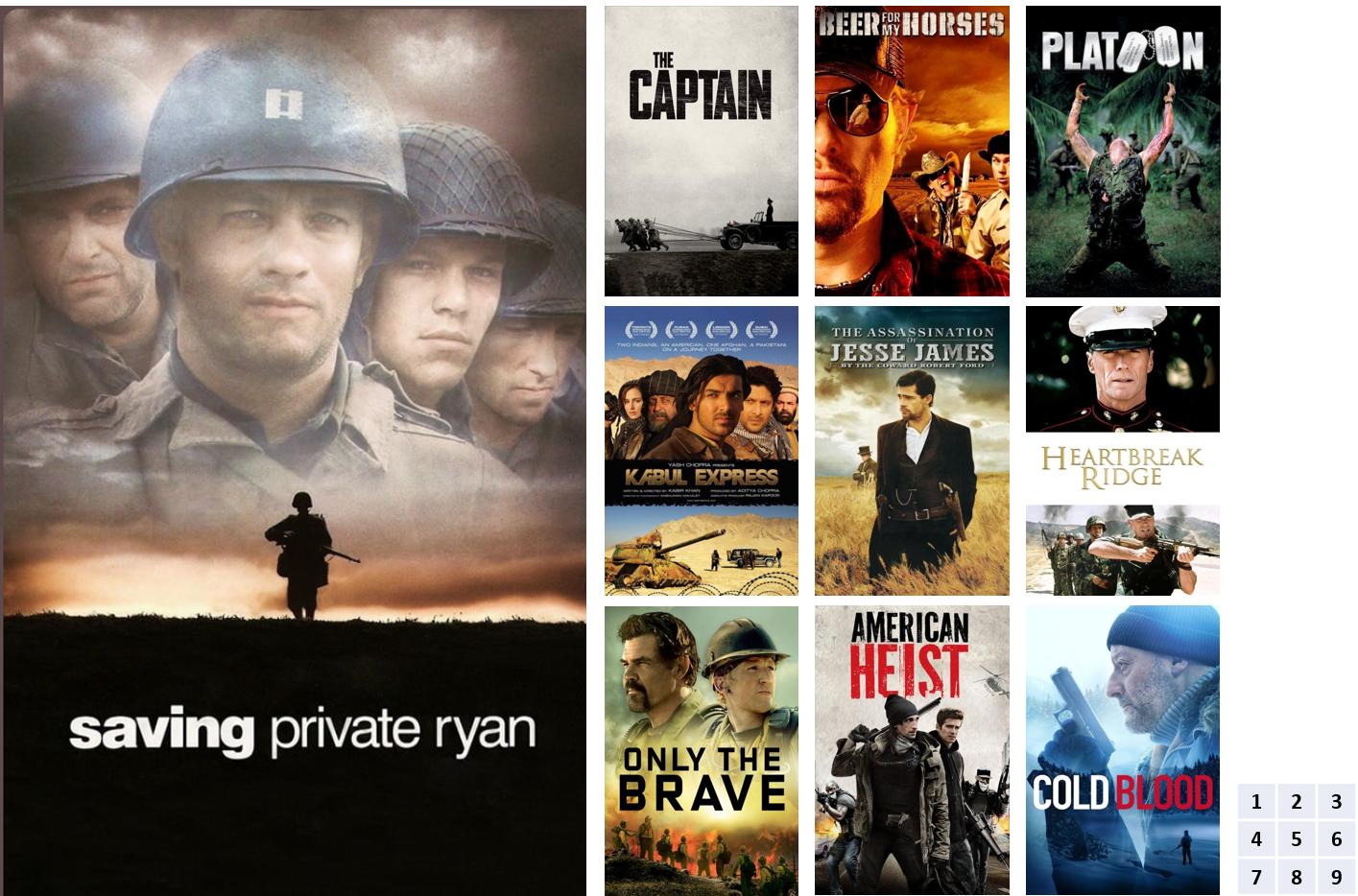}
    \caption{Poster retrieval using the keyword `war' in the context of a typical War movie \emph{Saving Private Ryan (1998)}}
\end{figure}

\begin{figure}[H]
    \centering
    \includegraphics[scale = 0.35]{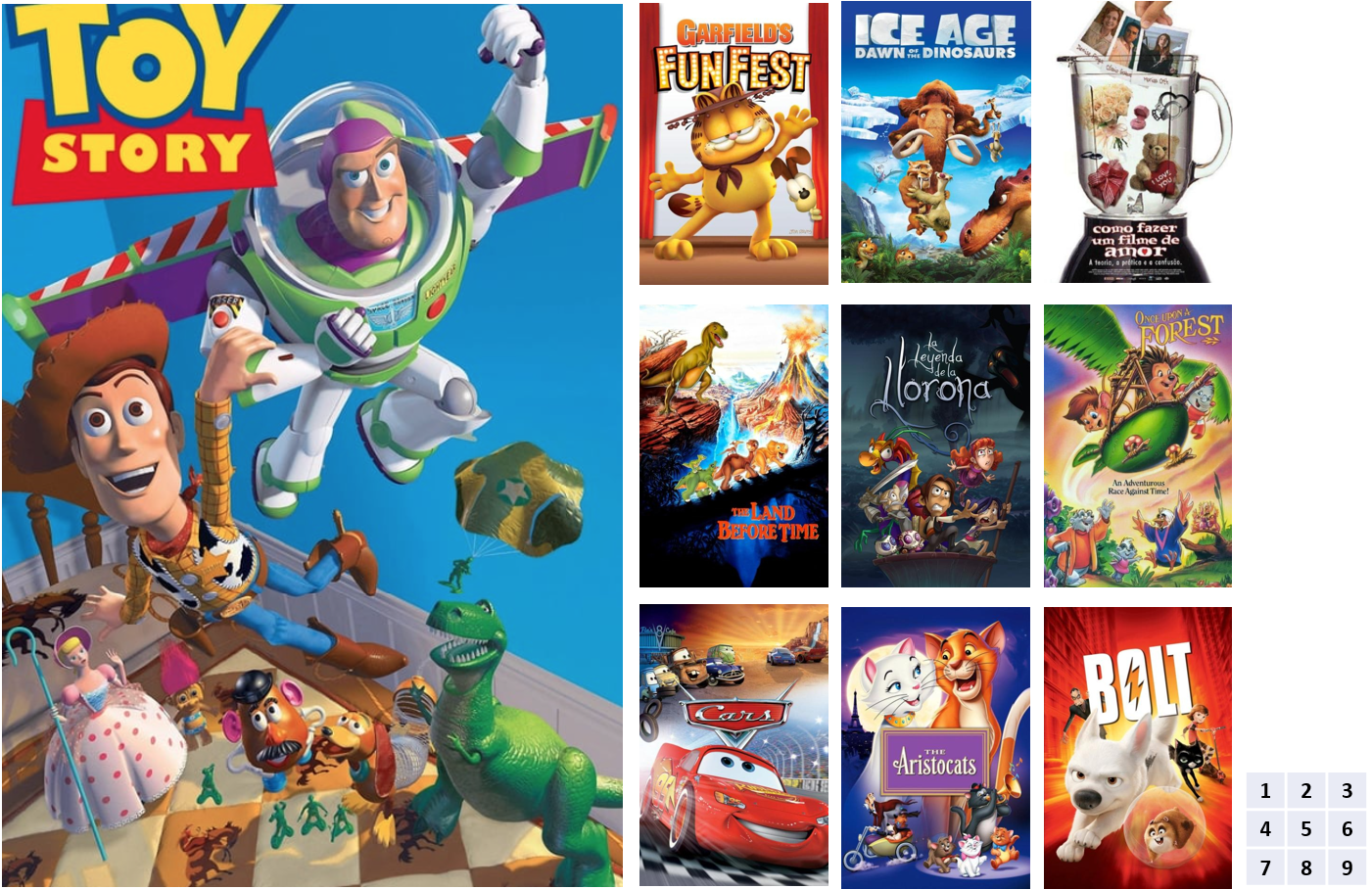}
    \caption{Poster retrieval using the keyword `friendship' in the context of a typical Family \& Animation movie \emph{Toy Story (1995)}}
\end{figure}

\end{document}